\documentclass[a4paper,hyper,published]{JHEP3}
\JHEP{03(2006)079}
\JHEPcopydate{2006}
\JHEPspecialurl{http://jhep.sissa.it/archive/papers/jhep032006079.pdf/jhep032006079.pdf}
\def\be{\begin{equation}}
\def\ee{\end{equation}}
\def\bea{\begin{eqnarray}}
\def\eea{\end{eqnarray}}
\def\ba{\begin{array}}
\def\ea{\end{array}}
\def\nn{\nonumber}

\def\trm{\textrm{Im}}
\title{Remarks on Hawking radiation as tunneling from the BTZ black holes}
\author{Shuang-Qing Wu and Qing-Quan Jiang \\ \small
College of Physical Science and Technology, Central China Normal University,
Wuhan, Hubei 430079, People's Republic of China \\
E-mail: \email{sqwu@phy.ccnu.edu.cn}, 
\email{jiangqingqua@126.com}}
\received{February 6, 2006} 
\accepted{March 13, 2006 \\ {\scshape Published}: March 24, 2006}

\abstract{Hawking radiation viewed as a semiclassical tunneling process from the event horizon of the
(2 + 1)-dimensional rotating BTZ black hole is carefully reexamined by taking into account not only the
energy conservation but also the conservation of angular momentum when the effect of the emitted particle's
self-gravitation is incorporated. In contrast to previous analysis of this issue in the literature, our
result obtained here fits well to the Kraus-Parikh-Wilczek's universal conclusion without any modification
to the Bekenstein-Hawking area-entropy formulae of the BTZ black hole. \\

}


\begin{document}

Ever since Parikh and Wilczek \cite{PW} put forward a semiclassical framework (see also \cite{KKW} for early
efforts underlying the basis of this framework, here we briefly refer it to as the Kruas-Parikh-Wilczek's
methodology) to implement Hawking radiation as a tunneling process across the horizon of the static spherically
symmetric black holes and demonstrated that the emission spectrum of black hole radiance is not strictly
pure thermal, there have been considerable efforts to generalize this work to those of various spherically
symmetric black holes \cite{PW,ETR1,ETR2}, and all the obtained results are very successful to support the
Kraus-Parikh-Wilczek's prescription \cite{PW,KKW}. There also have some recent attempts \cite{RBH} to extend
this approach to the case of some stationary axisymmetric geometries, however, to the best of our knowledge,
the treatments made in these researches \cite{RBH} are not completely satisfactory until very recently we
\cite{JW} present a plausible solution to this problem in the Kerr and Kerr-Newman black hole case.

As far as the case of a (2 + 1)-dimensional Ba\~{n}ados-Teitelboim-Zanelli (BTZ) \cite{BTZ} black hole is
concerned, several investigations along this line have also been done \cite{TRBTZ,ECV1,LWB,ECV2,AO}. 
For instance, Hawking radiation as a tunneling process from the charged and uncharged nonrotating BTZ black
holes was discussed in ref. \cite{TRBTZ}, while the spinning BTZ black hole case was investigated in
\cite{ECV1,LWB}. Vagenas \cite{ECV2} had also dealt with the case of the Ach\'{u}carro-Ortiz black hole
\cite{AO} which is the Kaluza-Klein reduction of the (2 + 1)-dimensional BTZ model. We mention that a
different methodology to apply the tunneling picture in the rotating BTZ black hole case has also been
advised \cite{ANVZ}, which may be attributed to the complex path method \cite{CPM}.

In this note, we shall reexamine Hawking radiation as tunneling from the rotating BTZ black hole by taking
into account not only the energy conservation but also the conservation of angular momentum and show that
the Kraus-Parikh-Wilczek's universal relation can be preserved without modifying the Bekenstein-Hawking entropy
expression of the BTZ black hole. This result is in contrast with previous analysis on the same subject \cite{ECV1},
where the main result obtained there does not fit well to the universal one given in refs. \cite{PW,KKW}, but
acquires a `self-gravitation' correction to the Bekenstein-Hawking entropy formulae. We think the reason for
this is that the rotation degree of freedom of the BTZ black hole had not been considered there. In addition,
we shall revisit the same subject in a different choice of coordinate gauge also.

The BTZ black hole solution is an exact solution to Einstein field equation in a (2 + 1)-dimensional theory
of gravity with a negative cosmological constant ($\Lambda = -1/l^2$). The corresponding line element is
\be
ds^2 = -\Delta dt^2 +\frac{dr^2}{\Delta} +r^2\Big(d\phi -\frac{J}{2r^2}dt\Big)^2 \, ,
\label{BTZm}
\ee
where the lapse function
\be
\Delta = -M +\frac{r^2}{l^2} +\frac{J^2}{4r^2} \, ,
\ee
in which two integration constants $M$ and $J$ denote the ADM mass and angular momentum of the BTZ black hole,
respectively. (The BTZ unit $G_3 = 1/8$ is adopted through out this paper.)

The outer (event) and inner horizons are given by the condition $\Delta = 0$, and read
\be
r_{\pm}^2 = l^2\frac{M\pm\sqrt{M^2 -J^2/l^2}}{2} \, .
\ee
The Bekenstein-Hawking entropy of the spinning BTZ black holes is twice the perimeter $L$ of the event horizon
\be
S(M, J, l) = 2L = 4\pi r_+ \, ,
\ee
the surface gravity and angular velocity at the event (outer) horizon are easily evaluated
\be
\kappa = \frac{1}{2}\frac{d\Delta}{dr}\Big|_{r = r_+} = \frac{r_+^2 -r_-^2}{r_+l^2} \, , \qquad
\Omega_+ = -\frac{g_{t\phi}}{g_{\phi\phi}}\Big|_{r = r_+} = \frac{J}{2r_+^2} = \frac{r_-}{r_+l} \, .
\ee

The idea of Kraus-Parikh-Wilczek's method is to describe the BTZ black hole background as dynamical by treating the
Hawking radiation as a tunneling process. In order to apply this methodology to the case of a rotating BTZ black
hole and to do an explicit computation of the tunneling rate, a key trick is to introduce a Painlev\'{e}-type
coordinate system that is well-behaved at the event (outer) horizon. There are essentially two slightly different
approaches to arrive at this aim. The first is that one must firstly find a Painlev\'{e}-BTZ coordinate system
and then transforms it into the dragging coordinate system. This method is well addressed by us in ref. \cite{JW},
where the reason why one must adopt a dragging coordinate system is also explained in details there. Vagenas
\cite{ECV1} only considered the energy conservation but no angular momentum conservation, i.e., he viewed the
ADM mass of the BTZ black hole can fluctuate but the angular momentum $J$ is kept fixed. [Note that the total
ADM mass is kept fixed while the mass of the BTZ black hole decreases due to the emitted radiation.] He only
performed the first step without further transforming it into the dragging coordinate system in his analysis;
at a cost, he had to modify the well-known Bekenstein-Hawking area-entropy formulae of the BTZ black hole. The
second one is to reverse the order of the above two steps. That is, one firstly transforms the metric (\ref{BTZm})
into the dragging coordinate system and then seeks a Painlev\'{e}-type coordinate transformation. In the former
step of this method, the resultant metric \cite{LWB,ECV2} for the spinning BTZ black hole is (after performing
a dragging coordinate transformation $d\phi = J/(2r^2)dt$)
\be
ds^2 = -\Delta dt^2 +\frac{dr^2}{\Delta} \, ,
\label{AOm}
\ee
which represents a (1 + 1)-dimensional hypersurface in the (2 + 1)-dimensional BTZ spacetime. This line
element (\ref{AOm}) is, in fact, identified with that of the Ach\'{u}carro-Ortiz black hole \cite{AO} which
is the Kaluza-Klein reduction of the (2 + 1)-dimensional BTZ black hole. The subsequent tunneling analysis
\cite{LWB} essentially reproduces those of a two-dimensional dilatonic black hole \cite{ECV2}.

It should be pointed out that the second method is only effective in the 3-dimensional case unlike the first
one which is universal in any dimensions. In what follows, we shall adopt the first method. For this purpose,
we introduce the time coordinate $\tau$ and the angular coordinate $\varphi $ by imposing the ans\"{a}tz:
\bea
dt &=& d\tau -\frac{\sqrt{1 -F}}{\Delta}dr \, , \label{tt} \\
d\phi &=& d\varphi -\Big(\frac{J}{2r^2}\Big)\frac{\sqrt{1 -F}}{\Delta}dr \label{pt} \, ,
\eea
where we choose the function $F = \Delta$ as Vagenas did in ref. \cite{ECV1}.

Obviously the above transformations (\ref{tt}) and (\ref{pt}) eliminate the coordinate singularities at the
horizons $r_{\pm}$. The line element (\ref{BTZm}) is now written as a Painlev\'{e}-type form
\bea
ds^2 &=& -\Delta d\tau^{2} +2\sqrt{1 -\Delta}d\tau dr +dr^2 +r^2\Big(d\varphi -\frac{J}{2r^2}d\tau\Big)^2 \nn \\
&=& -d\tau^{2} +\big(\sqrt{1 -\Delta}d\tau +dr\big)^2 +r^2\Big(d\varphi -\frac{J}{2r^2}d\tau\Big)^2 \, .
\label{PBTZ}
\eea

As explained in ref. \cite{JW}, there are two essential reasons that one must further transform the above
Painlev\'{e}-BTZ metric (\ref{PBTZ}) into a dragging coordinate system. The first reason is that just as the
original BTZ line element (\ref{BTZm}) the infinite red-shift surface $r_{TLS} = l\sqrt{M}$ does not coincide
with the outer horizon $r_+$ so that the geometrical optical limit cannot be used since the tunneling computation
is essentially a kind of WKB (s-wave) approximation. The other is that one must consider the dragging effect in
a rotating black hole spacetime since the physical field must be dragged also in a rotating background spacetime.
So a physically reasonable description must be within a dragging coordinate system, which can be equivalently
to state that an observer is rest at such a reference system. Moreover, by performing a dragging coordinate
transformation $d\varphi = J/(2r^2)d\tau$, the infinite red-shift surface coincides with the event horizon
in the new line element which shall be referred to as a dragged Painlev\'{e}-BTZ metric
\be
d\hat{s}^2 = -\Delta d\tau^{2} +2\sqrt{1 -\Delta}d\tau dr +dr^2
 = -d\tau^{2} +\big(\sqrt{1 -\Delta}d\tau +dr\big)^2 \, ,
\label{dPBTZ}
\ee
so that the WKB approximation can be applied now.

It is easily to observe that the above metric (\ref{dPBTZ}) can also be deduced from the line element of a
Ach\'{u}carro-Ortiz black hole \cite{AO} by carrying out only the temporal coordinate transformation (\ref{tt}).
This implies that the dragged Painlev\'{e}-BTZ (\ref{dPBTZ}) can also be obtained by means of the second method
mentioned above. The reason why two different procedures lead to the same result is that the transformations
(\ref{tt}) and (\ref{pt}) keep the dragging angular velocity $\Omega$ unchanged
\be
\Omega = \frac{d\phi}{dt} = \frac{d\varphi}{d\tau} = \frac{J}{2r^2} \, .
\ee
However it should be stressed that this coincidence only occurs in the three-dimensional case, in higher dimensions
two different methods in general derive two distinct line elements.

To apply the Kraus-Parikh-Wilczek's semiclassical tunneling analysis, the radial null ($d\hat{s}^2 = 0$) geodesics
followed by a massless particle need to be determined as follows
\be
\dot{r} \equiv \frac{dr}{d\tau} = \pm 1 -\sqrt{1 -\Delta} \, ,
\ee
where the signs $+$ and $-$ correspond to the outgoing and ingoing geodesics, respectively, under the assumption
that $\tau$ increases towards future.

Let us now focus on a semiclassical treatment of the associated radiation (outgoing massless particle). We adopt
the picture of a pair of virtual particles spontaneously created just inside the event horizon. The positive energy
particle tunnels out from the outer horizon while the negative partner is absorbed by the black hole resulting in
a decrease in the mass. In the case of the rotating BTZ black hole with fixed angular momentum $J$, the emitted
particle is simply visualized as a shell of energy (mass) $\omega$. Taking into account the energy conservation,
we must fix the total ADM mass and let the ADM mass $M$ of the BTZ black hole vary. If a shell of energy $\omega$
is radiated outwards the outer horizon, then the BTZ black hole mass will be reduced to $M -\omega$, so the line
element will be modified to
\be
d\hat{s}^2 = -\widetilde{\Delta}d\tau^{2} +2\sqrt{1 -\widetilde{\Delta}}d\tau dr +dr^2 \, ,
\label{mm1}
\ee
and the shell will accordingly travel on the modified geodesics
\be
\dot{r} = 1 -\sqrt{1 -\widetilde{\Delta}} = \frac{\widetilde{\Delta}}{1 +\sqrt{1 -\widetilde{\Delta}}} \, ,
\label{MRNG}
\ee
where the lapse function $\Delta$ is modified to $\widetilde{\Delta}$ with the replacement of mass $M$ by
$M -\omega$,
\be
\widetilde{\Delta} = -(M -\omega) +\frac{r^2}{l^2} +\frac{J^2}{4r^2} \, .
\ee

Since the emission rate from a spinning BTZ black hole can be expressed in terms of the imaginary part of the
action for an outgoing positive-energy particle as
\be
\Gamma = e^{-2\trm S} \, ,
\ee
we therefore need to evaluate the imaginary part of the action
\be
\trm S = \trm \int_{r_{in}}^{r_{out}}p_rdr = \trm \int^{r_{out}}_{r_{in}}\int_0^{p_r}dp_r^{\prime}dr\, ,
\ee
for such a particle which crosses the event horizon outwards from $r_{in}$ to $r_{out}$
(Note that: $r_{in} = r_+ > r_{out} = \tilde{r}_+$), where
\be
r_{out}^2 = l^2\frac{(M -\omega) +\sqrt{(M -\omega)^2 -J^2/l^2}}{2} \, .
\ee

The transition from the momentum variable to the energy variable can be made by using Hamilton's equation of
motion
\be
\dot{r} = \frac{dH}{dp_r} = \frac{d(M -\omega)}{dp_r} \, ,
\ee
and eq. (\ref{MRNG}). After some calculations we get the explicit result
\be
\trm S = \trm \int^{r_{out}}_{r_{in}}\int^{M -\omega}_M\frac{d(M -\omega^{\prime})dr}{1
-\sqrt{1 +(M -\omega^{\prime}) -\frac{r^2}{l^2} -{\frac{J^2}{4r^2}}}} = 2\pi(r_{in} -r_{out}) \, .
\ee
Apparently the emission probability depends not only on the mass $M$ and angular momentum $J$ of the BTZ
black hole but also on the energy $\omega$ of the emitted massless particle
\be
\Gamma = e^{-2\trm S} = e^{4\pi(r_{out} -r_{in})} = e^{\Delta S} \, ,
\ee
where $\Delta S = S(M -\omega, J, l) -S(M, J, l)$ is the change in the entropy of the BTZ black hole before
and after the emission of the shell of energy $\omega$.

We see that the tunneling rate fits well to the universal result obtained in Refs. \cite{PW,KKW}, without
modifying the standard area-entropy expression of a rotating BTZ black hole. This is in contrast with the
previous result obtained for a spinning BTZ black hole with a fixed $J$ \cite{ECV2}. We deduce that the
entropy of the (2 + 1)-dimensional BTZ black hole is still proportional to its horizon area, therefore,
it is clear that previous modification to the Bekenstein-Hawking area-entropy formulae for the BTZ black
hole is not necessary \cite{LWB} although the emitted particle's self-gravitation effect is incorporated.

However, it should be noted that the preceding discussion is limited to the case of energy conservation only.
Since we are considering a spinning BTZ black hole, so the rotation degree of freedom should be well addressed
also. This can be compensated by treating the emitted massless particle as a shell of energy $\omega$ and angular
momentum $j$. Now taking into account not only the energy conservation but also the conservation of angular
momentum, we must fix the total ADM mass and total angular momentum, but let the ADM mass $M$ and angular momentum
$J$ of the BTZ black hole vary. If a shell of energy $\omega$ and angular momentum $j$ tunnels out from the outer
horizon, then the BTZ black hole mass and angular momentum will be reduced to $M -\omega$ and $J -j$, respectively.
Therefore the modified line element and modified radial null geodesics are still respectively represented by eqs.
(\ref{mm1}) and (\ref{MRNG}) with the lapse function $\Delta$ replaced by
\be
\widetilde{\Delta} = -(M -\omega) +\frac{r^2}{l^2} +\frac{(J -j)^2}{4r^2} \, .
\ee

It follows that the evaluation of the imaginary part of the action is of no difficulty to complete by eliminating
a cyclic coordinate $\varphi$ in the Lagrangian function and goes as follows
\bea
\trm S &=& \trm \Big[\int_{r_{in}}^{r_{out}}p_rdr -\int_{\varphi_{in}}^{\varphi_{out}}p_{\varphi}d\varphi\Big]
= \trm \int^{r_{out}}_{r_{in}}\Big[\int_0^{p_r}dp_r^{\prime}dr
-\int_0^{p_{\varphi}}dp_{\varphi}^{\prime}d\varphi\Big] \nn \\
&=& \trm \int^{r_{out}}_{r_{in}}\Bigg[\int_{(0, 0)}^{(p_{r}, p_{\varphi})}\Big(\dot{r}dp_r^{\prime}
-\dot{\varphi}dp_{\varphi}^{\prime}\Big)\Bigg]\frac{dr}{\dot{r}} \, ,
\label{IA}
\eea
where $p_r$ and $p_{\varphi}$ are two canonical momenta conjugate to the coordinates $r$ and $\varphi$, respectively.
As before, $r_{in} = r_+$ and $r_{out} = \tilde{r}_+$, which is given by
\be
r_{out}^2 = l^2\frac{(M -\omega) +\sqrt{(M -\omega)^2 -(J -j)^2/l^2}}{2} = \tilde{r}_+^2 \, ,
\ee
are the locations of the event horizon before and after a particle tunnels out.

To remove the momentum in favor of energy, we can make use of the Hamilton's equations
\bea
\dot{r} &=& \frac{dH}{dp_r}\Big|_{(r; \varphi, p_{\varphi})}
 = \frac{d(M -\omega)}{dp_r} \, , \nn \\
\dot{\varphi} &=& \frac{dH}{dp_{\varphi}}\Big|_{(\varphi; r, p_r)}
 = \widetilde{\Omega}\frac{d(J -j)}{dp_{\varphi}} \, ,
 \label{HE}
\eea
where $dH_{(\varphi; r, p_r)} = \widetilde{\Omega}d(J -j)$ represents the energy change of the black hole
because of the loss of angular momentum when a particle tunnels out, and the dragging angular velocity is
given by
\be
\widetilde{\Omega} = \frac{J -j}{2r^2} \, .
\ee

Now substituting eqs. (\ref{MRNG}) and (\ref{HE}) into (\ref{IA}) and interchanging the order of integration
yields
\bea
\trm S &=& \trm \int^{r_{out}}_{r_{in}}\int_{(M, J)}^{(M -\omega, J -j)}\Big[d(M -\omega^{\prime})
-\widetilde{\Omega}d(J -j^{\prime})\Big]\frac{dr}{\dot{r}} \nn \\
&=& \trm \int_{(M, J)}^{(M -\omega, J -j)}\int^{r_{out}}_{r_{in}}
\frac{1 +\sqrt{1 -\widetilde{\Delta}}}{\widetilde{\Delta}}\Big[d(M -\omega^{\prime})
-\frac{(J -j^{\prime})}{2r^2}d(J -j^{\prime})\Big]dr \nn \\
&=& 2\pi(r_{in} -r_{out}) \, ,
\eea
thus the tunneling probability
\be
\Gamma = e^{-2\trm S} = e^{4\pi(r_{out} -r_{in})} = e^{\Delta S} \, ,
\ee
reproduces the Kraus-Parikh-Wilczek's standard result for the tunneling picture, where $\Delta S = S(M -\omega,
J -j, l) -S(M, J, l)$ is the change in the entropy of the BTZ black hole before and after the particle emission.
Again we see that there is no need to modify the well-known Bekenstein-Hawking area-entropy relation of a spinning
BTZ black hole even when we incorporate the particle's self-gravitation effect.

In what follows, we would like to point out that the ans\"{a}tz for the coordinate transformations (\ref{tt})
and (\ref{pt}) with the choice of the function $F = \Delta$ is not unique. Alternatively we can make a different
choice of the function
\be
F = \frac{\Delta}{\Delta_0} \, , \qquad \Delta_0 = \Delta\big|_{M = 0} = \frac{r^2}{l^2} +\frac{J^2}{4r^2} \, .
\ee
With this choice for the coordinate transformations (\ref{tt}, \ref{pt}), the line element (\ref{BTZm}) is now
transformed into the form
\bea
ds^2 &=& -\Delta d\tau^{2} +2\sqrt{1 -\frac{\Delta}{\Delta_0}}d\tau dr +\frac{dr^2}{\Delta_0}
+r^2\Big(d\varphi -\frac{J}{2r^2}d\tau\Big)^2 \nn \\
&=& -\Delta_0 d\tau^{2} +\Big(\sqrt{\Delta_0 -\Delta}d\tau +\frac{dr}{\sqrt{\Delta_0}}\Big)^2
+r^2\Big(d\varphi -\frac{J}{2r^2}d\tau\Big)^2 \, .
\eea
Similarly the dragging coordinate transformation $d\varphi = J/(2r^2)d\tau$ further reduces the above line element
to
\be
d\hat{s}^2 = -\Delta d\tau^{2} +2\sqrt{1 -\frac{\Delta}{\Delta_0}}d\tau dr +\frac{dr^2}{\Delta_0}
 = -\Delta_0 d\tau^{2} +\frac{1}{\Delta_0}\Big(\sqrt{1 -\frac{\Delta}{\Delta_0}}d\tau +dr\Big)^2 \, ,
\label{mm2}
\ee
from which the radial null ($d\hat{s}^2 = 0$) geodesics followed by a massless particle are determined
\be
\dot{r} = \Delta_0\Big(\pm 1 -\sqrt{1 -\Delta/\Delta_0}\Big) \, .
\ee
Incidentally we point out that the line element (\ref{mm2}) can also be derived by using the second approach
mentioned above.

Now the explicit tunneling computation completely repeats the preceding procedure with the emitted particle being
viewed as a shell of energy $\omega$ and angular momentum $j$. By considering the particle's self-gravitation
effect and the conservation of energy as well as that of angular momentum, when a shell of energy $\omega$ and
angular momentum $j$ is radiated outwards the outer horizon, then the ADM mass $M$ and angular momentum $J$ of
the BTZ black hole will be reduced to $M -\omega$ and $J -j$, respectively. Accordingly the line element and
radial null geodesics will be modified as
\bea
&& d\hat{s}^2 = -\widetilde{\Delta}d\tau^{2} +2\sqrt{1 -\widetilde{\Delta}/\widetilde{\Delta}_0}d\tau dr
+\frac{dr^2}{\widetilde{\Delta}_0} \, , \\
&& \dot{r} = \widetilde{\Delta}_0\Big(1 -\sqrt{1 -\widetilde{\Delta}/\widetilde{\Delta}_0}\Big)
= \frac{\widetilde{\Delta}}{1 +\sqrt{1 -\widetilde{\Delta}/\widetilde{\Delta}_0}} \, ,
\eea
where
\be
\widetilde{\Delta} = -(M -\omega) +\frac{r^2}{l^2} +\frac{(J -j)^2}{4r^2} \, , \qquad
\widetilde{\Delta}_0 = \frac{r^2}{l^2} +\frac{(J -j)^2}{4r^2} \, .
\ee

Note that near the outer horizon $r \approx \tilde{r}_+$, the modified radial null geodesic has the asymptotic
behavior
\be
\dot{r} \approx \tilde{\kappa}(r -\tilde{r}_+) \, , \qquad
\tilde{\kappa} = \frac{1}{2}\frac{d\widetilde{\Delta}}{dr}\Big|_{r = \tilde{r}_+} \, ,
\ee
and the dragging angular velocity takes its value at the event horizon
\be
\widetilde{\Omega}_+ = \frac{J -j}{2\tilde{r}_+^2} \, ,
\ee
so it is easily to see that the imaginary part of the action is dominated by its value at the outer horizon
\bea
\trm S &=& \trm \int^{r_{out}}_{r_{in}}\int_{(M, J)}^{(M -\omega, J -j)}\Big[d(M -\omega^{\prime})
-\widetilde{\Omega}d(J -j^{\prime})\Big]\frac{dr}{\dot{r}} \nn \\
&\approx& \trm \int_{(M, J)}^{(M -\omega, J -j)}\int^{r_{out}}_{r_{in}}\Big[d(M -\omega^{\prime})
-\widetilde{\Omega}_+d(J -j^{\prime})\Big]\frac{dr}{\tilde{\kappa}(r -\tilde{r}_+)} \nn \\
&=& -\int_{(M, J)}^{(M -\omega, J -j)}\frac{\pi}{\tilde{\kappa}}\Big[d(M -\omega^{\prime})
-\frac{(J -j^{\prime})}{2\tilde{r}_+^2}d(J -j^{\prime})\Big] \, .
\label{IAI}
\eea
On the other hand, from the event horizon equation
\be
\widetilde{\Delta}\big|_{r = \tilde{r}_+} = -(M -\omega) +\frac{\tilde{r}_+^2}{l^2}
+\frac{(J -j)^2}{4\tilde{r}_+^2} = 0 \, ,
\ee
one can use a simple algebra approach to prove the differential mass formulae \cite{FLBTZ} of the first
law of a spinning BTZ black hole
\be
d(M -\omega) -\frac{(J -j)}{2\tilde{r}_+^2}d(J -j)
 = \frac{d\widetilde{\Delta}}{dr}\Big|_{r = \tilde{r}_+} d\tilde{r}_+
 = 2\tilde{\kappa} d\tilde{r}_+ \, .
\label{DFL}
\ee
Inserting the identity (\ref{DFL}) into the integral (\ref{IAI}), the final result is very concise
\be
\trm S = -2\pi\int^{r_{out}}_{r_{in}}d\tilde{r}_+^{\prime} = 2\pi(r_{in} -r_{out}) = -\frac{1}{2}\Delta S \, ,
\ee
and restores the standard expression for the tunneling rate.

To summarize, we have revisited Hawking radiation as a semiclassical tunneling process from the outer horizon of
a rotating BTZ black hole by taking into account the energy conservation and the conservation of angular momentum.
The Kraus-Parikh-Wilczek's universal conclusion can be retained without modifying the well-known Bekenstein-Hawking
area-entropy formulae of the BTZ black hole. This result is in contrast with previous analysis of the same subject
in the literature. Besides, we have also shown that the so-called Painlev\'{e}-type coordinate transformation is
not unique, two different choices of the coordinate transformation lead to the same result. This demonstrates that
our discussion present here is consistent on its behalf.

\section*{Acknowledgements}

S.-Q.Wu was supported by a starting fund from Central China Normal University and by the Natural Science
Foundation of China.


\end{document}